\documentclass[twocolumn,preprintnumbers,amsmath,amssymb,floatfix, superscriptaddress]{revtex4-2}

\usepackage{graphicx}

\usepackage{array}
\usepackage{dcolumn}
\usepackage{bm,epsfig}
\usepackage{epstopdf}
\usepackage[percent]{overpic}
\usepackage{physics,xfrac}
\usepackage{mathtools}
\setcitestyle{super}
\usepackage{color}
\usepackage{float}

\begin{document}

\title{Robust narrow-gap semiconducting behavior in square-net La$_{3}$Cd$_{2}$As$_{6}$}

\author{Mario M. Piva}
\email{Mario.Piva@cpfs.mpg.de}
\affiliation{Los Alamos National Laboratory, Los Alamos, New Mexico 87545, USA}
\affiliation{Instituto de F\'{\i}sica ``Gleb Wataghin'', UNICAMP, 13083-859, Campinas, SP, Brazil}
\affiliation{Max Planck Institute for Chemical Physics of Solids, N\"{o}thnitzer Str.\ 40, D-01187 Dresden, Germany}

\author{Marein C. Rahn}
\affiliation{Los Alamos National Laboratory, Los Alamos, New Mexico 87545, USA}
\affiliation{Institute for Solid State and Materials Physics, Technical University of Dresden, 01062 Dresden, Germany}

\author{Sean M. Thomas}
\affiliation{Los Alamos National Laboratory, Los Alamos, New Mexico 87545, USA}

\author{Brian L. Scott}
\affiliation{Los Alamos National Laboratory, Los Alamos, New Mexico 87545, USA}

\author{Pascoal G. Pagliuso}
\affiliation{Instituto de F\'{\i}sica ``Gleb Wataghin'', UNICAMP, 13083-859, Campinas, SP, Brazil}

\author{Joe D. Thompson}
\affiliation{Los Alamos National Laboratory, Los Alamos, New Mexico 87545, USA}

\author{Leslie M. Schoop}
\affiliation{Department of Chemistry, Princeton University, Princeton, NJ, 08544, USA}

\author{Filip Ronning}
\affiliation{Los Alamos National Laboratory, Los Alamos, New Mexico 87545, USA}

\author{Priscila F. S. Rosa}
\affiliation{Los Alamos National Laboratory, Los Alamos, New Mexico 87545, USA}

\date{\today}

\begin{abstract}

ABSTRACT: Narrow-gap semiconductors are sought-after materials due to their potential for long-wavelength detectors, thermoelectrics, and more recently non-trivial topology. Here we report the synthesis and characterization of 
a new family of narrow-gap semiconductors, $R$$_{3}$Cd$_{2}$As$_{6}$ ($R=$ La, Ce). Single crystal 
x-ray diffraction at room temperature reveals that the As square nets distort and Cd vacancies order in a monoclinic superstructure. A putative charge-density ordered state sets in at 279~K in La$_{3}$Cd$_{2}$As$_{6}$ and at 136~K in Ce$_{3}$Cd$_{2}$As$_{6}$ 
and is accompanied by a substantial increase in the electrical resistivity in both compounds. The resistivity of the La member increases by thirteen orders of magnitude on cooling, which points to a remarkably clean semiconducting ground state. Our results suggest that light square net materials within a $I4/mmm$ parent structure are promising clean narrow-gap semiconductors. 

\centering
\includegraphics{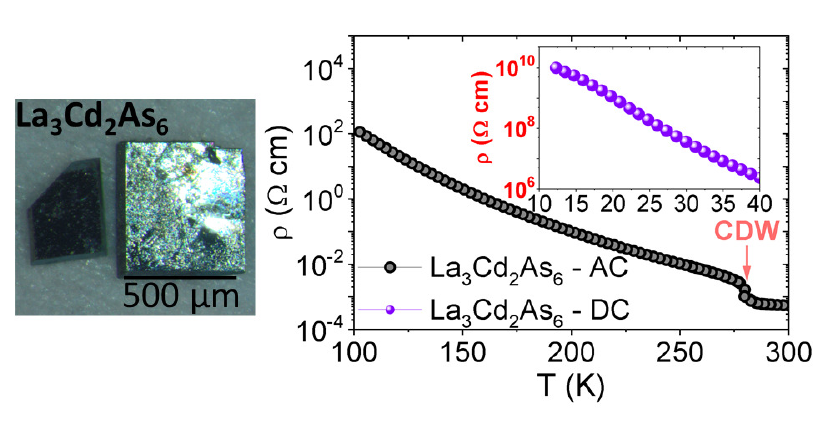}

\end{abstract}

\maketitle

\section{INTRODUCTION}

Electronic instabilities are ubiquitous in quantum materials. Notable examples 
include spin-density waves, superconductivity, and charge-density waves (CDWs). 
The latter is a modulation of conduction electron density
accompanied by a periodic distortion of the crystal lattice \cite{ReviewCDW,ReviewCDW2,CDWlowD2}.
Interestingly, CDWs have been recently observed in close proximity
to exotic correlated phenomena, such as high-temperature superconductivity and axionic topological states \cite{CDW,cuprates2019,axions2019,axions2021}.

In addition, the formation of a CDW phase minimizes
the energy of the system by opening a gap 
at the Fermi level, which may drive the material towards a 
narrow-gap semiconducting state \cite{ReviewCDW,ReviewCDW2,CDWlowD2}. Disorder-free narrow-gap semiconducting materials are of interest due to their potential functionalities, 
which include thermoelectricity, dark-matter detection, and nontrivial topology \cite{Thermo,DarkM,PbTeThermo}. Their narrow gap, however, often suffers from conducting in-gap states created by defects or impurities \cite{Book,Book2}.

One way to design narrow-gap semiconductors is 
to search for low-dimensional crystal structures, which are fundamentally unstable to the formation
of energy gaps from small distortions \cite{CDWlowD2,ReviewCDW,ReviewCDW2}. 
In fact, CDW phases were initially identified in one-dimensional (1D) transition metal 
trichalcogenides $MX_{3}$ ($M =$ Nb and Ta, $X =$ S, Se, or Te) containing chains of NbSe$_{6}$ prisms 
and in quasi-2D layered transition metal 
dichalcogenides \cite{ReviewCDW,ReviewCDW2,CDW}. 

Here we consider layered tetragonal materials with the ``112" general formula
$RMX_{2}$ ($R =$ lanthanide, $M =$ Au, Ag, Cu, Cd, and Zn, $X =$ Bi, Sb, and As), which contain pnictide square nets. 
Previous band structure analysis and molecular orbital models have shown that square nets built from
more electronegative elements are prone to distortions that lift the tetragonal symmetry \cite{Sqrnets_0}.
This tendency occurs for two reasons. First, the $R$-square net $p$-bands close to the Fermi level ($E_{F}$) and the $M$ $d$ bands become well separated, which allows electron transfer and the formation of anionic units, that form covalent bonds leading to Zintl phases \cite{Zintl_1,Zintl_2,Zintl_Review,Zintl_Rosa}. Second, there is greater mixing between $s$ and $p$ orbitals in pnictide elements with small atomic number ($Z$), which (again) favors the classic
octet rule in a distorted structure over hypervalent bonds in an undistorted square net \cite{Sqrnets_1,Sqrnets_2}.

Important insights into distorted structures come from chalcogenide-based square nets, and numerous polytellurides host CDW phases with distorted Te square nets. $RE$Te$_{3}$ ($RE$ = rare-earth) \cite{ReTe3,SmTe3,CeTe3}, Pb$_{3-x}$Sb$_{1+x}$S$_{4}$Te$_{2 - \delta}$ \cite{PbSbSTe}, Sm$_{2}$Te$_{5}$ \cite{Sm2Te5}, Cu$_{0.63}$EuTe$_{2}$ \cite{CuEuTe}, $AMR$Te$_{4}$ ($A$ = K, Na and $M$ = Cu, Ag) \cite{AMRETe4}, and K$_{1/3}$Ba$_{2/3}$AgTe$_{2}$ \cite{KBaAgTe2} are a few examples. In the latter, cation sites are partially occupied in order to satisfy the Zintl rule in the distorted structure. Importantly, the primary cause of the observed vacancy superstructure is argued to be a distortion in the tellurium square net, rather than the partial or complete ordering of the cations \cite{KBaAgTe2}. Yet many distorted polytellurides exhibit metallic behavior \cite{RTe3metal,SmTemetal,CuEuTe,AMRETe4}. Recently, GdTe$_{3}$ has been shown to display a remarkably high electron mobility of 61,200~cm$^{2}$V$^{-1}$s$^{-1}$ at low temperatures \cite{GdTe3}. Ultrahigh mobilities ($>$ 10,000~cm$^{2}$V$^{-1}$s$^{-1}$) and an anomalous Hall effect are also observed in pnictide square net materials such as EuMnBi$_{2}$ \cite{EuMnBi2}.

\begin{figure}[!t]
\includegraphics[width=0.48\textwidth]{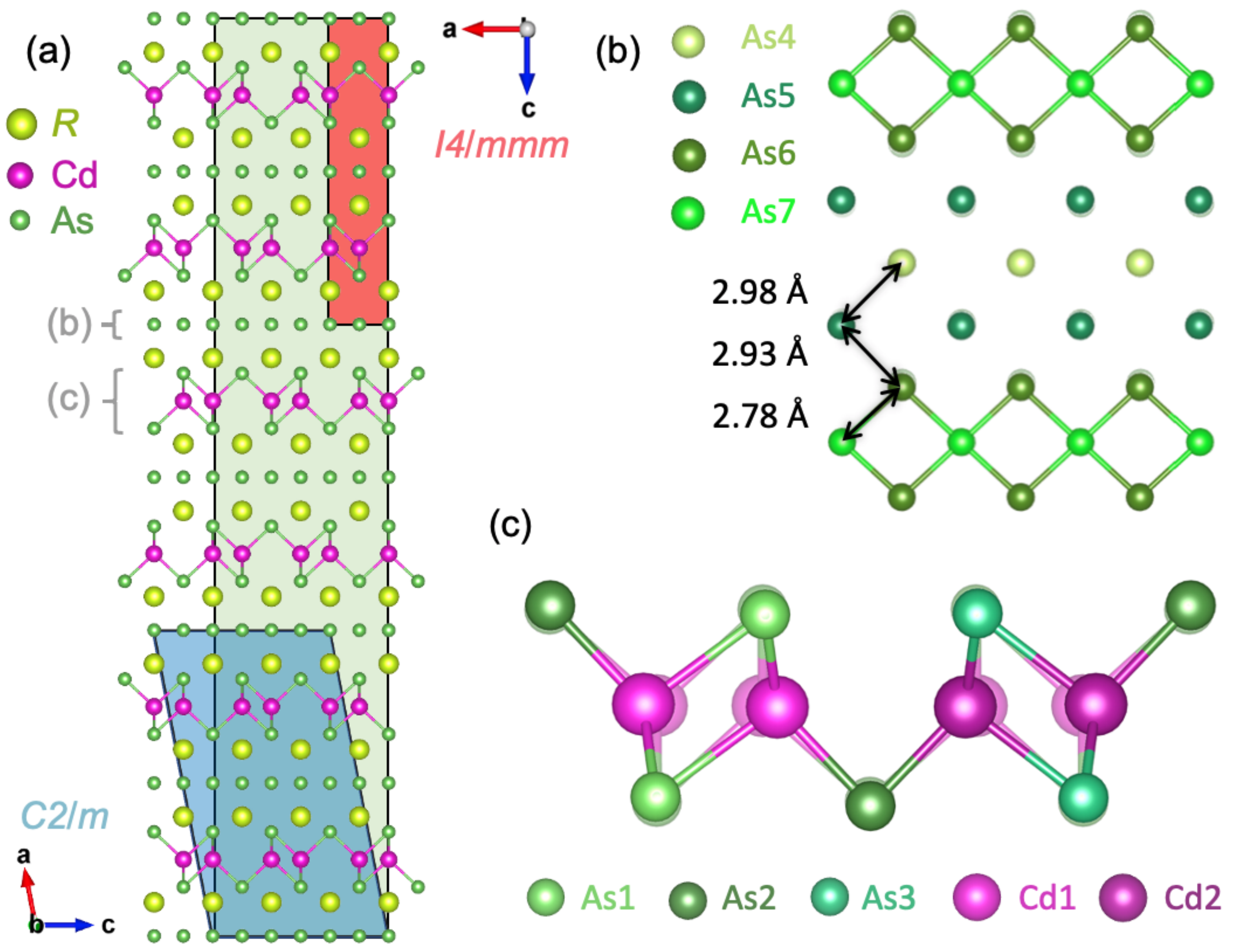}
\caption{(a) Structural relationship of the tetragonal CeCdAs$_2$ parent ($I4/mmm$, red), the conceptual orthorhombic 3$\times$1$\times$3 vacancy superstructure (green) and the true, monoclinic, Ce$_3$Cd$_2$As$_6$ unit cell ($C2/m$, blue). (b) Detailed view of the As square-net layers. To emphasize the stripe distortion, only the short As6--As7 bonds are drawn. (c) Detailed view of the CdAs layers. For reference, the nominal undistorted structure is drawn as a shaded background. Cd ions are significantly displaced towards the vacancy stripes. As a consequence, the apical As ions As1 and As3 are slightly drawn towards the Cd layer, while the As2 ions dangling above/below the Cd-vacancy stripes are slightly pushed away from the Cd layer. }
\label{Fig1}
\end{figure}

Here we highlight a route for 
clean narrow-gap semiconducting behavior in distorted arsenic square nets. We investigate a rather underexplored phase space in which $M=$ post-transition metal \cite{layeredAs1,layeredAs2}. The 
substantial electronegativity of low-$Z$ arsenic combined with the decrease in metallic character of post-transition metals
provides a promising route to the realization of distortion-driven charge-density waves.

In this paper, we report the synthesis and characterization of distorted LaCd$_{2/3}$As$_{2}$ and CeCd$_{2/3}$As$_{2}$. 
At room temperature, both compounds crystallize in a vacancy superstructure of the $I4/mmm$ ``112" parent
structure and show semiconducting behavior. The ordered superstructure thus corresponds to the ``326" stoichiometry. As temperature is decreased, both compounds display
a phase transition at $T_{\mathrm{CDW}} = $ 136~K and 278~K for Ce$_{3}$Cd$_{2}$As$_{6}$ and La$_{3}$Cd$_{2}$As$_{6}$, respectively. Given the many structural degrees of freedom due to the broken tetragonal symmetry, we attribute these transitions to CDW-driven structural distortions. This picture also explains the opening of a gap in the electronic density of states. Remarkably, the electrical 
resistivity of La$_{3}$Cd$_{2}$As$_{6}$ increases by thirteen orders of magnitude on cooling below $T_{\mathrm{CDW}}$, in agreement 
with the vanishingly small Sommerfeld coefficient from 
specific heat measurements. The estimated activation energies from Arrhenius plots are
105(1)~meV and 74(1)~meV for La$_{3}$Cd$_{2}$As$_{6}$ and Ce$_{3}$Cd$_{2}$As$_{6}$, respectively. 

 \begin{figure*}[!t]
	\includegraphics[width=\textwidth]{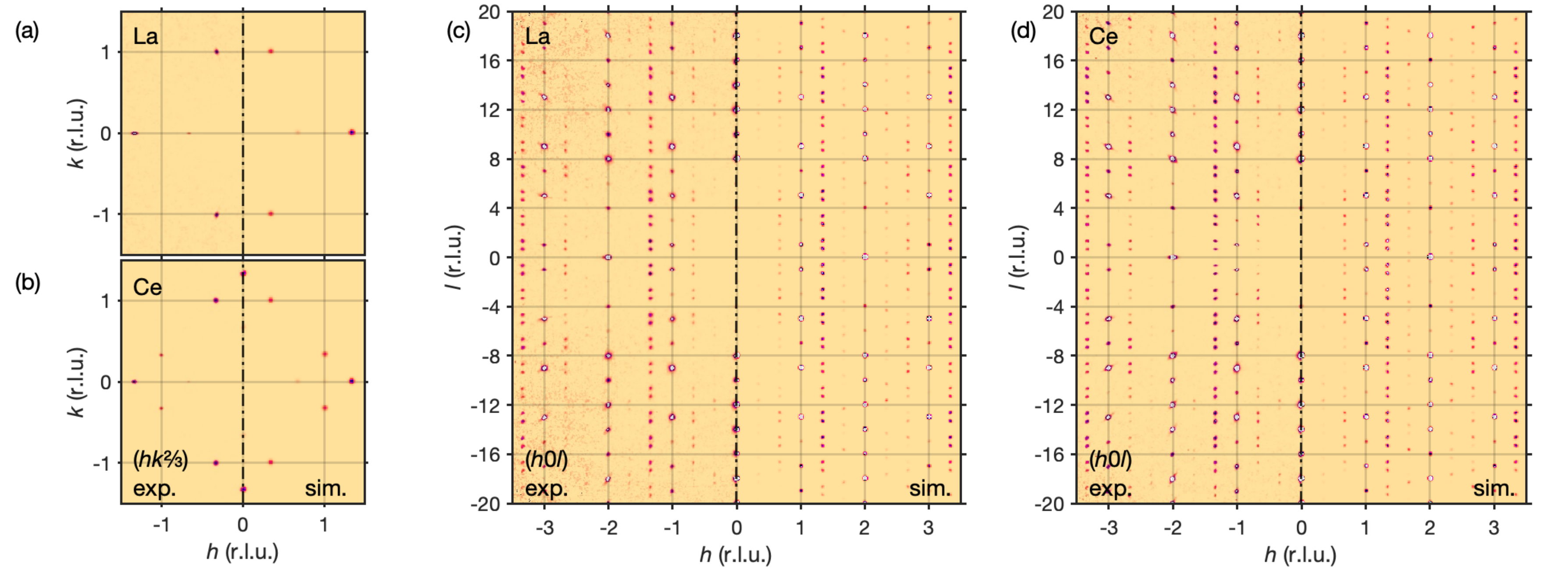}%
	\caption{\label{Fig2} Comparison of experimental and calculated x-ray diffraction intensity maps of reciprocal space. The left and right halves of each panel show the measured and calculated intensity, respectively. For convenience, the data is indexed in the $I4/mmm$ parent cell. (a,b) $(hk\frac23)$ slices reveal the presence of (a) two stripe domains (i.e., twins) in LaCd$_{0.67}$As$_2$ and (b) four in CeCd$_{0.67}$As$_2$. (c,d)  ($h0l$) slices show a large number of $q_\text{Cd}=\langle\frac23 0 \frac23\rangle$ superstructure peaks. The modulations of the intensities of these peaks is highly sensitive of small distortions of the ionic arrangement away from the nominal positions in the tetragonal parent.}
\end{figure*}

\section{EXPERIMENTAL DETAILS}

Black plate-like single crystals of Ce$_{3}$Cd$_{2}$As$_{6}$ and La$_{3}$Cd$_{2}$As$_{6}$ were grown by the vapor transport technique. 
First, a polycrystalline seed of 1La:0.7Cd:2As was prepared via solid state reaction at 800~$^{\circ}$C. 
Then the polycrystalline powder was loaded along with iodine in a quartz tube, which was sealed in vacuum. 
The tube was kept in a temperature gradient from 830~$^{\circ}$C to 720~$^{\circ}$C for a week. The initial polycrystalline material was kept in the hot zone, and single crystals precipitated in the cold zone. The synthesized phase was investigated by Mo $K_\alpha$ single crystal x-ray diffraction and energy-dispersive x-ray spectroscopy (EDX) both at room temperature. EDX measurements confirmed the 3:2:6 stoichiometry within experimental error. The specific heat of a collection of approximately 10 single crystals (total mass $\approx 1$~mg) was measured using a Quantum Design PPMS that employs a quasi-adiabatic thermal relaxation technique. The electrical resistivity ($\rho$) was characterized with the same instrument in the standard four-probe configuration.
The current was applied in the basal plane of the crystal. For Ce$_{3}$Cd$_{2}$As$_{6}$, an AC bridge was used to measure $\rho$ in all temperature range. While, for La$_{3}$Cd$_{2}$As$_{6}$, $\rho$ was measured with an AC bridge, at high temperatures, whereas at low temperatures a DC method was required due to the large resistance of the sample. 

\section{RESULTS}

\begin{table}[b]
	\begin{center}
		\begin{tabular}{ l c c c c c }
			\multicolumn{6}{l}{\textbf{$\boldsymbol{R}$Cd$_{2/3}$As$_{2}$}, $I4/mmm$ (\#139), $Z=4$ }\\
			\multicolumn{6}{l}{$a_t=b_t\sim4.1\,$\AA, $c_t\sim21.3\,$\AA}\\ [5pt]
			\multicolumn{2}{l}{ion, Wyck.} & $x$ & $y$ & $z$& occ. (\%)\\
			\hline
			$R$  & $4e$ &    $0$     &    $0$     &  0.11         & 100 \\
			Cd  & $4d$ &    $0$     & $\sfrac12$ & $\sfrac14$    & 66  \\ 
			As1 & $4c$ &    $0$     & $\sfrac12$ &    $0$        & 100 \\
			As2 & $4e$ &    $0$     &    $0$     &  0.34         & 100 \\
			\hline
		\end{tabular}
		\caption{ \label{Tab01} Structural parameters of $R_{3}$Cd$_{2}$As$_{6}$, inferred from refinements in a tetragonal cell (neglecting superstructure reflections). A detailed summary of refined parameters and uncertainties for either compound ($R=$ La, Ce) is provided in the Supporting Information \cite{sm}.}
	\end{center}
\end{table}

We start with the structural characterization of Ce$_{3}$Cd$_{2}$As$_{6}$ and La$_{3}$Cd$_{2}$As$_{6}$. For both materials, Bragg reflections can be indexed in a tetragonal unit cell corresponding to the respective parent compounds with ``112'' stoichiometry. The approximate parameters of this $I4/mmm$ structure are given in Table~\ref{Tab01}. The corresponding tetragonal cell is shaded red in Fig.~\ref{Fig1}(a). A detailed summary of the refined parameters of either compound is provided in the Supporting Information \cite{sm}. The occupancy of the Cd site ($4d$ in $I4/mmm$) converges close to 66\% in both materials, confirming the stoichiometry measured in EDX.

Interestingly, the Cd vacancies are not randomly distributed, but rather order in a stripe pattern. Fig.~\ref{Fig2} shows x-ray intensity maps interpolated to the $(h,k,\frac23)$ and $(h0l)$ planes of reciprocal space (with reference to the $I4/mmm$ cell). Aside from the integer-index Bragg peaks, a pattern of weak satellite reflections is observed, which is described by the propagation vector $\mathbf{q}_\mathrm{Cd}=\langle\frac23,0,\frac23\rangle$. This indicates an ordering of the vacancies in a 3$\times$1$\times$3 superstructure [green cell in \ref{Fig1}(a)], which can be reduced to a base-centered monoclinic structure of space group $C2/m$ [blue cell in \ref{Fig1}(a)]. 

\begin{table}[b]
 \begin{center}
	\begin{tabular}{llllllll}
		\multicolumn{8}{l}{\textbf{\textit{R}$_3$Cd$_2$As$_6$}, $C2/m$ (\#12, unique axis $b$, cell choice 1)}\\
		\hline 
		\multicolumn{2}{l}{$\vec{a}=\vec{a}_t-\vec{c}_t$ } & ,~~ i.e.~~$a=\sqrt(a_t^2+c_t^2)$ & \multicolumn{5}{l}{ $\sim21.6$\,\AA  }    \\
		\multicolumn{2}{l}{$\vec{b}=-\vec{b}_t$ }          & ,~~ i.e.~~$b= a_t$              & \multicolumn{5}{l}{ $\sim4.1$\,\AA }       \\
		\multicolumn{2}{l}{$\vec{c}=-3\,\vec{a}_t$}        & ,~~ i.e.~~$c= 3\,a_t$            & \multicolumn{5}{l}{ $\sim12.2$\,\AA }   \\
		\multicolumn{8}{l}{$\alpha=\gamma=90^\circ$}\\
		\multicolumn{8}{l}{$\beta= 90^\circ+\arctan(a_t/c_t) \sim 100.8^\circ$ } \\[5pt]
	\end{tabular}
	\begin{tabular}{>{\centering\arraybackslash}lc>{\centering\arraybackslash}p{13mm}>{\centering\arraybackslash}p{7mm}>{\centering\arraybackslash}p{19mm}>{\centering\arraybackslash}p{12mm}>{\centering\arraybackslash}p{12mm}}
		&       & \multicolumn{3}{>{\centering\arraybackslash}p{39mm}}{nominal position} & \multicolumn{2}{>{\centering\arraybackslash}p{24mm}}{~~deviation (\AA)}\\
		\multicolumn{2}{l}{ ion Wyck.}  & $x$ & $y$ & $z$ & d$x$  & d$z$ \\
		\hline
		$R$1 & $4i$  & $z_{\mathrm{t,}R}$  &     0       & $z_{\mathrm{t,}R}/3$         &  ~~~      & ~~~        \\
		$R$2 & $4i$  & $z_{\mathrm{t,}R}$  &     0       & $(z_{\mathrm{t,}R}+2)/3$     &  ~~~      & ~~~        \\
		$R$3 & $4i$  & $z_{\mathrm{t,}R}$  &     0       & $(z_{\mathrm{t,}R}+1)/3$     &  ~~~      & ~~~        \\ 
		Cd1 & $4i$  &   $\sfrac14$     &     0       & $\sfrac{11}{12}$                 &  ~~~      & $-0.27$    \\
		Cd2 & $4i$  &   $\sfrac14$     &     0       & $\sfrac{7}{12}$                  &  ~~~      & $+0.23$     \\
		As1 & $4i$  & $z_\mathrm{t,As}$  &     0       & $z_\mathrm{t,As}/3$            &  $-0.05$  & ~~~        \\
		As2 & $4i$  & $z_\mathrm{t,As}$  &     0       & $(z_\mathrm{t,As}+2)/3$        &  $+0.11$  & ~~~        \\
		As3 & $4i$  & $z_\mathrm{t,As}$  &     0       & $(z_\mathrm{t,As}+1)/3$        &  $-0.05$  & ~~~        \\
		As4 & $2b$  &        0         & $\sfrac12$  &     0                            &  ~~~     & ~~~        \\
		As5 & $4i$  &        0         &     0       & $\sfrac16$                       &  ~~~      & $+0.13$    \\
		As6 & $4i$  &    $\sfrac12$    &     0       & $\sfrac13$                       &  ~~~      & $+0.19$    \\
		As7 & $2c$  &        0         &    0        & $\sfrac12$                       &  ~~~      & ~~~        \\
		\hline
	\end{tabular}
		\caption{ \label{Tab02} Description of $R_3$Cd$_2$As$_6$ in the monoclinic setting. The parameters $a_\mathrm{t}$, $c_\mathrm{t}$, $z_{\mathrm{t, }R}\sim0.11$ and $z_\mathrm{t, As}\sim0.34$ refer to the $I4/mmm$ parent cells, as stated in Table~\ref{Tab01}. All sites are fully occupied. The refined $x$ and $z$ coordinates of the Wyckoff site $4i$ for either compound are given in the Supporting Information \cite{sm}. The right column states the resulting approximate deviation (in \AA) with respect to the nominal positions in the tetragonal parent cell.}
	\end{center}
\end{table}

 \begin{figure*}[!t]
	\includegraphics[width=\textwidth]{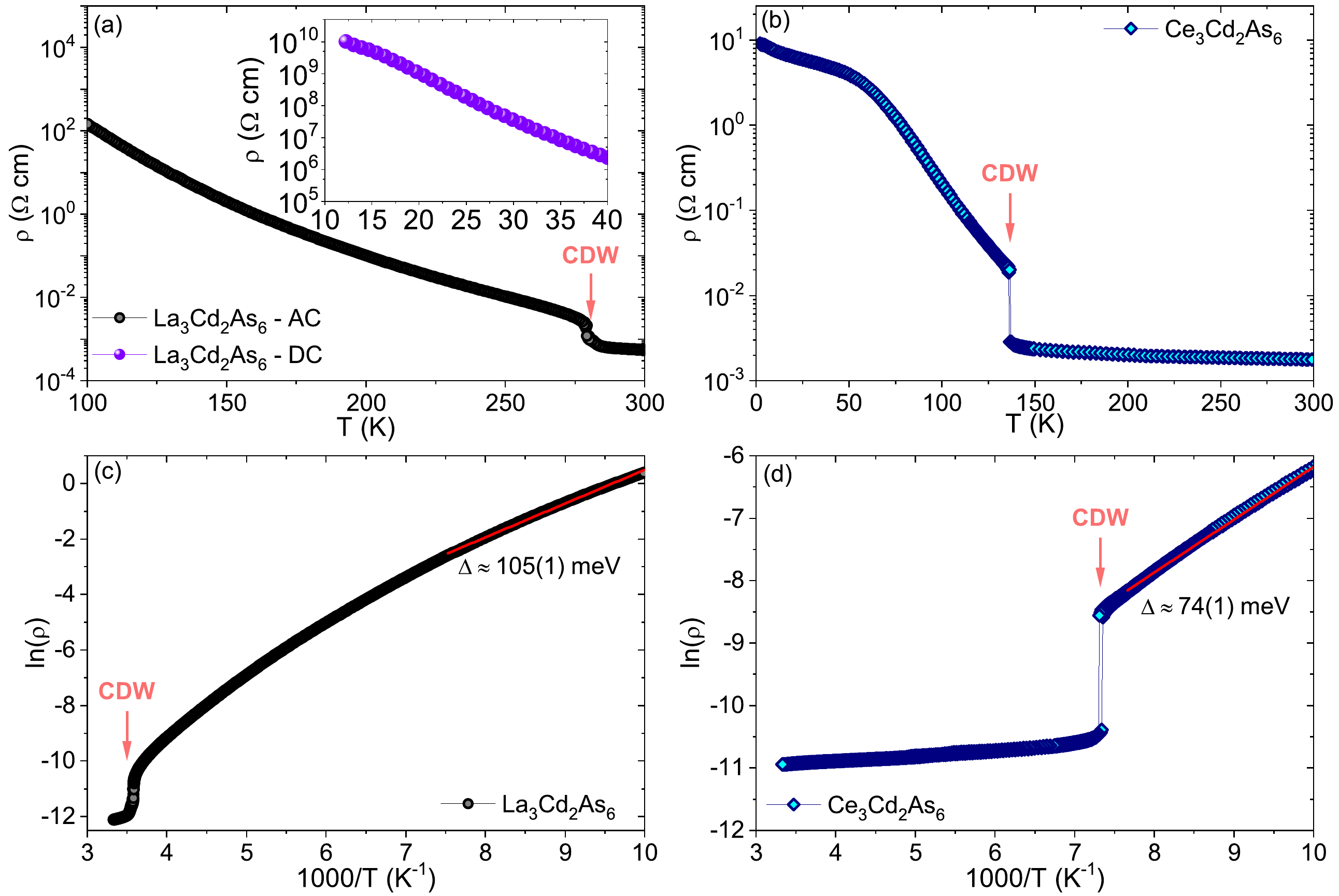}
	\caption{(a) Electrical resistivity as a function of temperature for La$_{3}$Cd$_{2}$As$_{6}$. The inset shows the electrical resistivity of La$_{3}$Cd$_{2}$As$_{6}$ in the  low temperature range. (b) Electrical resistivity as a function of temperature for Ce$_{3}$Cd$_{2}$As$_{6}$.   Natural logarithm of the electrical resistivity as a function of 1000 over the temperature for La$_{3}$Cd$_{2}$As$_{6}$ (c) and for Ce$_{3}$Cd$_{2}$As$_{6}$ (d). The solid red lines are linear fits.}
	\label{RxT}
\end{figure*}

Similar vacancy order has previously been observed in PrZn$_{0.67}$As$_2$ ($P4/nmm$). In that case, the vacancy pattern also results in a stoichiometric ``326'' compound, but with orthorhombic (\textit{Pmmn}) symmetry~\cite{Pr326}. By contrast, the vacancy stripes in Ce$_{3}$Cd$_{2}$As$_{6}$ and La$_{3}$Cd$_{2}$As$_{6}$ are staggered from one layer to the next in a sequence of six Cd layers. The transformation between the $I4/mmm$ and $C2/m$ cells, as illustrated in Fig.~\ref{Fig2}, is described in Table~\ref{Tab02}. The resulting staircase of Cd-vacancy stripes breaks the parent compounds' fourfold symmetry, which is accompanied by several structural modifications. While the $I4/mmm$ parent structure is determined by two ionic position parameters, twenty are required to define the $C2/m$ counterpart (cf. Table~\ref{Tab02}).

The satisfactory refinement within $I4/mmm$ when superstructure reflections are ignored shows that these distortions are weak. Nonetheless, small deviations of these parameters away from their value in the $I4/mmm$ setting cause significant variations of the structure factors of the superstructure peaks. This is illustrated in the reciprocal x-ray scattering intensity maps in Fig.~\ref{Fig2}. For comparison, each panel shows experimental intensities (left halves) along with our models (right halves). Details of this analysis are provided in the Supporting Information \cite{sm}. Major ionic displacements are summarized in Table~\ref{Tab02} in terms of the deviations (d$x$ and d$z$) relative to the ionic positions in the tetragonal parent compound. The full summary of refined parameters and uncertainties is given in the Supporting Information \cite{sm}.

The refinement adequately captures that the Bragg intensities at $(\frac230l)$ are significantly weaker than those along $(\frac430l)$. This is due to a ``relaxation'' of Cd ions by about $0.25$\,\AA~towards the vacancy stripes, as illustrated in Fig.~\ref{Fig1}(c). As expected, the displacement of Cd ions also leads to a measurable displacement of their ligands (As1, As2 and As3), which causes a characteristic modulation of superstructure intensities along the $l$ direction in reciprocal space. 

Importantly, the arsenic square nets in the new ``326'' structure are also distorted. Here, the high-symmetry As1 site of $I4/mmm$ splits into As4 -- As7 (in $C2/m$). The As-As bond-lengths are then modulated by about $0.2$~\AA, as illustrated in Fig.~\ref{Fig1}(b), and become shorter than hypervalent Sb-Sb distances. In the distorted case, the tolerance factor defined by $t = d_{\textmd{As-As}}/d_{\textmd{Ce/La-As}}$ is no longer valid \cite{Sqrnets_1,Sqrnets_2}. Whether the square-net distortion leads to Cd vacancies, similar to K$_{1/3}$Ba$_{2/3}$AgTe$_{2}$ in Ref. [22], or vice-versa remains an open question.

Finally, we note that no major structural differences have been observed between La$_{3}$Cd$_{2}$As$_{6}$ 
and Ce$_{3}$Cd$_{2}$As$_{6}$, aside from the expected lattice contraction from La to Ce, namely the volume
of the Ce unit cell being smaller by 2.1\,\%. Nonetheless, the La$_3$Cd$_2$As$_6$ crystal investigated 
featured only two of the four expected crystallographic twins corresponding to four propagation directions
of the vacancy staircase, as observed in Ce$_{3}$Cd$_{2}$As$_{6}$ [Figs.~\ref{Fig2}(a,b)]. This observation could stem from extrinsic reasons, such as finite strain during crystal growth.

We turn to the physical properties of La$_{3}$Cd$_{2}$As$_{6}$ single crystals. Figures~\ref{RxT}(a) and (b) display the in-plane electrical resistivity ($\rho$) of La$_{3}$Cd$_{2}$As$_{6}$ and Ce$_{3}$Cd$_{2}$As$_{6}$, respectively. Upon cooling, a sudden increase in the resistivity is observed at $T_{\mathrm{CDW}} = $ 279 K for La$3$Cd$2$As$6$ and at $T_{\mathrm{CDW}} = $ 136 K for Ce$_{3}$Cd$_{2}$As$_{6}$. Below $T_{\mathrm{CDW}}$, the observed semiconducting behavior indicates the opening of an energy gap due to the CDW phase transition, similar to Sr$_{3}$Ir$_{4}$Sn$_{13}$, Ca$_{3}$Ir$_{4}$Sn$_{13}$ \cite{CDWSC}, Ce$_{3}$Co$_{4}$Sn$_{13}$ and La$_{3}$Co$_{4}$Sn$_{13}$ \cite{3413synth,Ce3413-CDW,Ce3413-pressure,La3413-pressure}. Notably, 
the resistivity of La$_{3}$Cd$_{2}$As$_{6}$ increases up to 
10$^{10}$~$\Omega$\ cm at 12~K, as presented in the inset of Fig.~\ref{RxT}(a). 
This large increase of 
thirteen orders in magnitude is a rare property in narrow-gap semiconductors, which are typically prone to  impurity bands that may dominate the electronic properties at low temperatures \cite{Book,Book2}. Remarkably, this increase was achieved in an as-grown single crystal, which suggests that sample 
quality could be improved through further purification. The broad hump around 50~K for Ce$_{3}$Cd$_{2}$As$_{6}$ may be 
associated with the depopulation of the first excited crystal-field state. The magnetic properties of Ce$_{3}$Cd$_{2}$As$_{6}$ will be the focus of a separate study.

Figures~\ref{RxT}(c) and (d) present the resistivity in $\Omega$~m on a logarithmic scale, as a function of $(1000/T)$ for La$_{3}$Cd$_{2}$As$_{6}$ and Ce$_{3}$Cd$_{2}$As$_{6}$. An activated gap appears to open for temperatures below the CDW transition temperatures in both compounds.  The activation energies can be estimated from linear fits, considering the Arrhenius equation 
ln$(\rho_{xx}) =\ $ln$(\rho_{0}) + \Delta/T$, yielding 105(1)~meV and 74(1)~meV for La$_{3}$Cd$_{2}$As$_{6}$ and Ce$_{3}$Cd$_{2}$As$_{6}$, respectively. We note that the gap values are an estimation and should be taken with caution. Further, it is clear from Fig.~\ref{RxT}(c) that the gap value changes as a function of temperature. A temperature dependent activation energy may be caused by changes in the lattice parameters, due to thermal contraction, as observed in  Ce$_{3}$Bi$_{4}$Pt$_{3}$ \cite{Thermal}. Moreover, the presence of even small amounts of defects create extrinsic channels of conduction, such as variable range hopping and in-gap states \cite{Book,Book2}, which can also be a sensible scenario for a temperature dependent gap. Optical measurements are needed to directly probe the semiconducting gap of La$_{3}$Cd$_{2}$As$_{6}$ and Ce$_{3}$Cd$_{2}$As$_{6}$. 

\begin{figure}[!t]
	\includegraphics[width=0.48\textwidth]{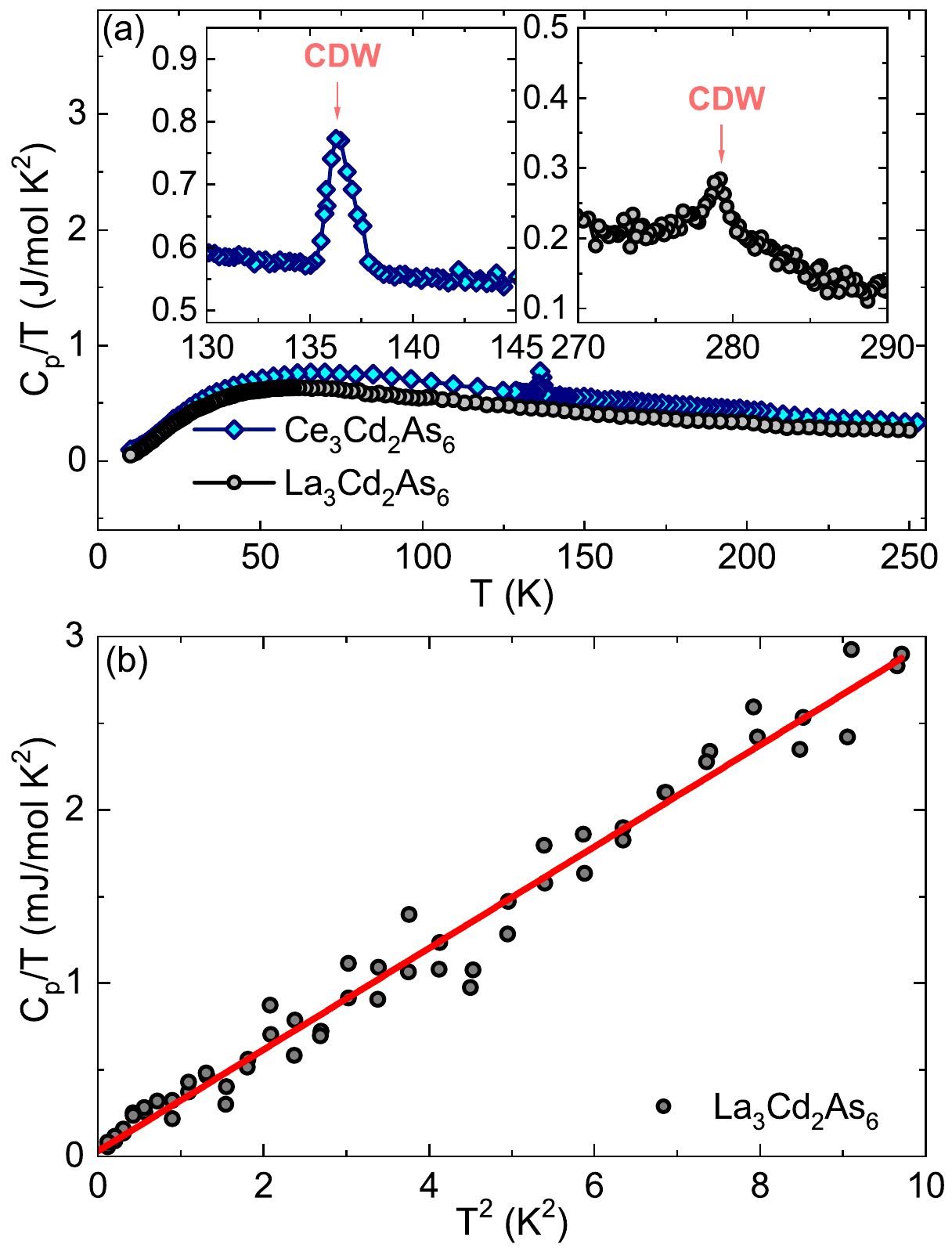}
	\caption{(a) Specific heat over the temperature ($C_{p}/T$) as a function of temperature for Ce$_{3}$Cd$_{2}$As$_{6}$ and La$_{3}$Cd$_{2}$As$_{6}$. The insets show a zoom in view of the CDW phase transitions. (b) $C_{p}/T$ as a function of $T^{2}$ for La$_{3}$Cd$_{2}$As$_{6}$ The solid red line is a linear fit.}
	\label{cp}
\end{figure}

Figure~\ref{cp}(a) displays the specific heat per mole of lanthanide divided by temperature ($C_{p}/T$) for both Ce$_{3}$Cd$_{2}$As$_{6}$ and La$_{3}$Cd$_{2}$As$_{6}$ 
as a function of temperature.  The insets of Fig.~\ref{cp}(a) present a detailed view of the transition temperatures, 
which are characterized by peaks at 279~K and 136~K for La$_{3}$Cd$_{2}$As$_{6}$ and Ce$_{3}$Cd$_{2}$As$_{6}$, respectively, in 
agreement with the electrical resistivity measurements. At this temperature scale, the relevance of magnetic Ce correlations to this transition is unlikely. Therefore, given the structural degree of freedom of As ions in the monoclinic structure, we attribute these transitions to a CDW phase caused by a further distortion in the arsenic square net. Figure~\ref{cp}(b) displays $C_{p}/T$ per mole of lanthanide as a function of $T^{2}$ for La$_{3}$Cd$_{2}$As$_{6}$. 
The solid red line is the linear fit used to estimate the Sommerfeld coefficient 
($\gamma$) of 0.03(3)~mJ/molK$^{2}$ and a Debye temperature ($\theta_{\mathrm{D}}$) of 291(2)~K for La$_{3}$Cd$_{2}$As$_{6}$. A negligible value of $\gamma$ is expected in clean semiconducting compounds 
and confirms the high quality of La$_{3}$Cd$_{2}$As$_{6}$ single crystals.

The strong suppression of the CDW transition temperature and activation energy going from La$_{3}$Cd$_{2}$As$_{6}$ to Ce$_{3}$Cd$_{2}$As$_{6}$ is an indication that the application of external pressure may be an effective tuning parameter, given the volume unit cell contraction by about 2~\% in Ce$_{3}$Cd$_{2}$As$_{6}$ as compared to La$_{3}$Cd$_{2}$As$_{6}$. High pressure experiments are needed to investigate whether the CDW phases in La$_{3}$Cd$_{2}$As$_{6}$ and Ce$_{3}$Cd$_{2}$As$_{6}$ can be suppressed, which may give rise to a superconducting state as in Sr$_{3}$Ir$_{4}$Sn$_{13}$, Ca$_{3}$Ir$_{4}$Sn$_{13}$ \cite{CDWSC} and La$_{3}$Co$_{4}$Sn$_{13}$ \cite{La3413-pressure}. Moreover, chemical pressure may also be an effective tuning parameter, and hence single crystals with smaller lanthanide elements, such as Pr or Nd, are desirable. Finally, spectroscopic experiments are necessary to search for electronic instabilities, which could be driving the CDW phase in La$_{3}$Cd$_{2}$As$_{6}$ and Ce$_{3}$Cd$_{2}$As$_{6}$.

\section{CONCLUSIONS}

In summary, we report the structural and electrical transport properties of narrow-gap 
semiconductors Ce$_{3}$Cd$_{2}$As$_{6}$ and La$_{3}$Cd$_{2}$As$_{6}$, which crystallize 
in a distorted variant of the $I4/mmm$ tetragonal structure $R$Cd$_{2/3}$As$_{2}$. The resulting base-centered monoclinic superstructure with $C2/m$ symmetry displays ordered Cd vacancies in a stripe pattern and a distorted As square net. 
Notably, both compounds feature charge density wave 
phases below 136~K (Ce$_{3}$Cd$_{2}$As$_{6}$) and 279~K (La$_{3}$Cd$_{2}$As$_{6}$), which suggests further distortion in the As square net. 
The CDW phases in these materials 
create gaps with estimated values of 105(1)~meV and 74(1)~meV for La$_{3}$Cd$_{2}$As$_{6}$ and 
Ce$_{3}$Cd$_{2}$As$_{6}$. A remarkable increase of thirteen orders of 
magnitude in the electrical resistivity was found for La$_{3}$Cd$_{2}$As$_{6}$ upon cooling from 
room temperature to 12~K. Coupled to the vanishingly small specific heat Sommerfeld coefficient, this result not only points to a robust semiconducting ground state in this class of compounds but also provides a route to realize clean narrow-gap semiconductors in distorted arsenic square net materials. 

\section{Supporting Information}

The Supporting Information consists of tables summarizing the crystallographic structure data for La$_{3}$Cd$_{2}$As$_{6}$ and Ce$_{3}$Cd$_{2}$As$_{6}$ and the complete crystallographic information files for both compounds at room temperature.  

\begin{acknowledgments}
We acknowledge useful discussions with Peter Abbamonte. Work at Los Alamos was performed under the auspices of the U.S. Department of Energy, Office of Basic Energy Sciences, Division of Materials Science and Engineering under project ``Quantum Fluctuations in Narrow-Band Systems". This work was also supported by the S\~ao Paulo Research Foundation (FAPESP) grants 2015/15665-3, 2017/10581-1, 2017/25269-3, CNPq grant $\#$ 304496/2017-0 and CAPES, Brazil. Scanning electron microscope and energy dispersive X-ray measurements were performed at the Center for Integrated Nanotechnologies, an Office of Science User Facility operated for the U.S. Department of Energy (DOE) Office of Science. This research has been supported by the Deutsche Forschungsgemeinschaft through the SFB 1143 and the Würzburg-Dresden Cluster of Excellence EXC 2147 (ct.qmat). Work at Princeton was supported by the Arnold and Mabel Beckman Foundation through a Beckman Young Investigator grant awarded to L.M.S.
\end{acknowledgments}

\bibliography{basename of .bib file}

\begin{figure*}[!t]
\hspace*{-1.2in}
\vspace*{-1.5in}
\centering
\includegraphics[width=1.3\textwidth]{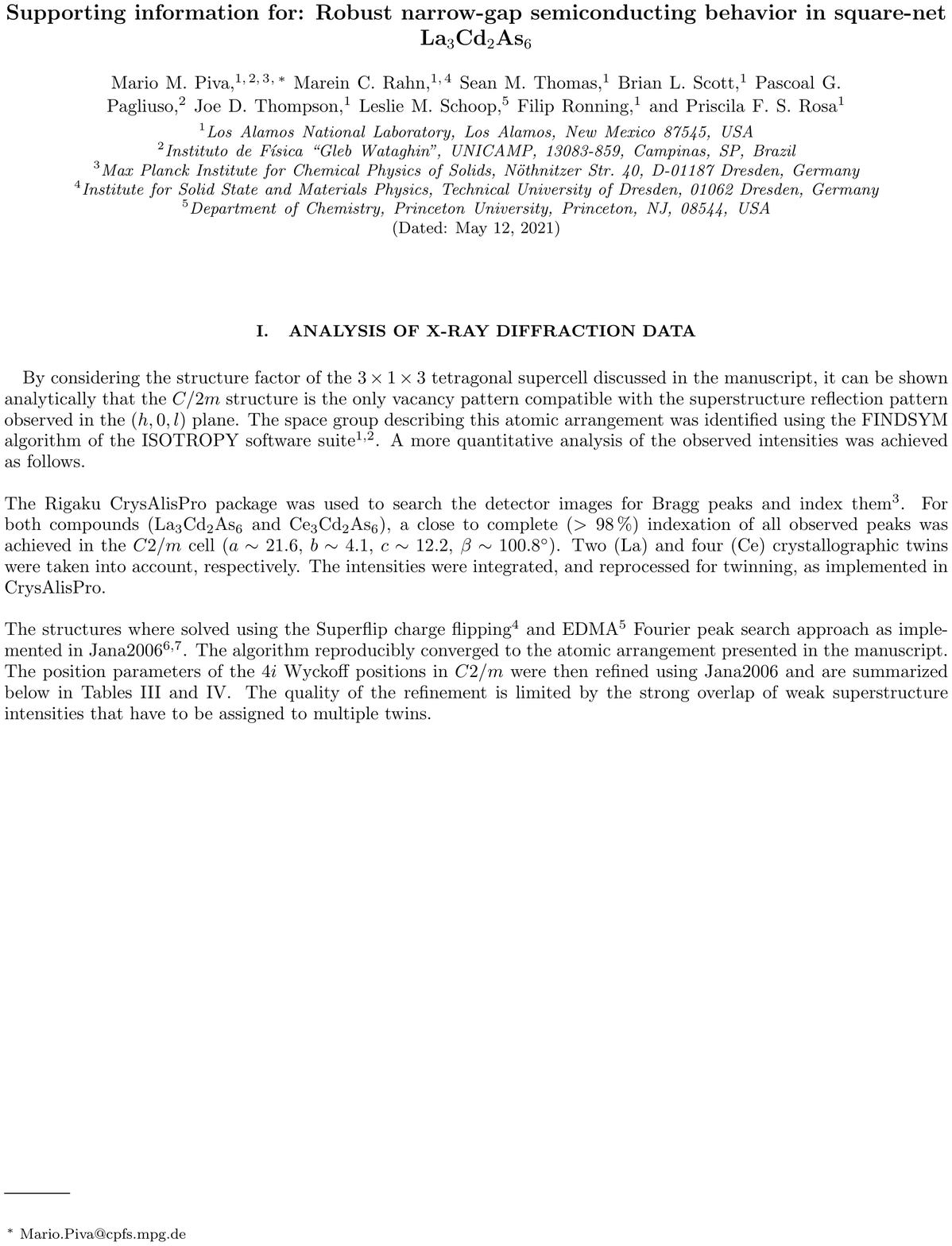}
\end{figure*}

\begin{figure*}[!t]
	\hspace*{-1.2in}
	\vspace*{-1.5in}
	\centering
	\includegraphics[width=1.3\textwidth]{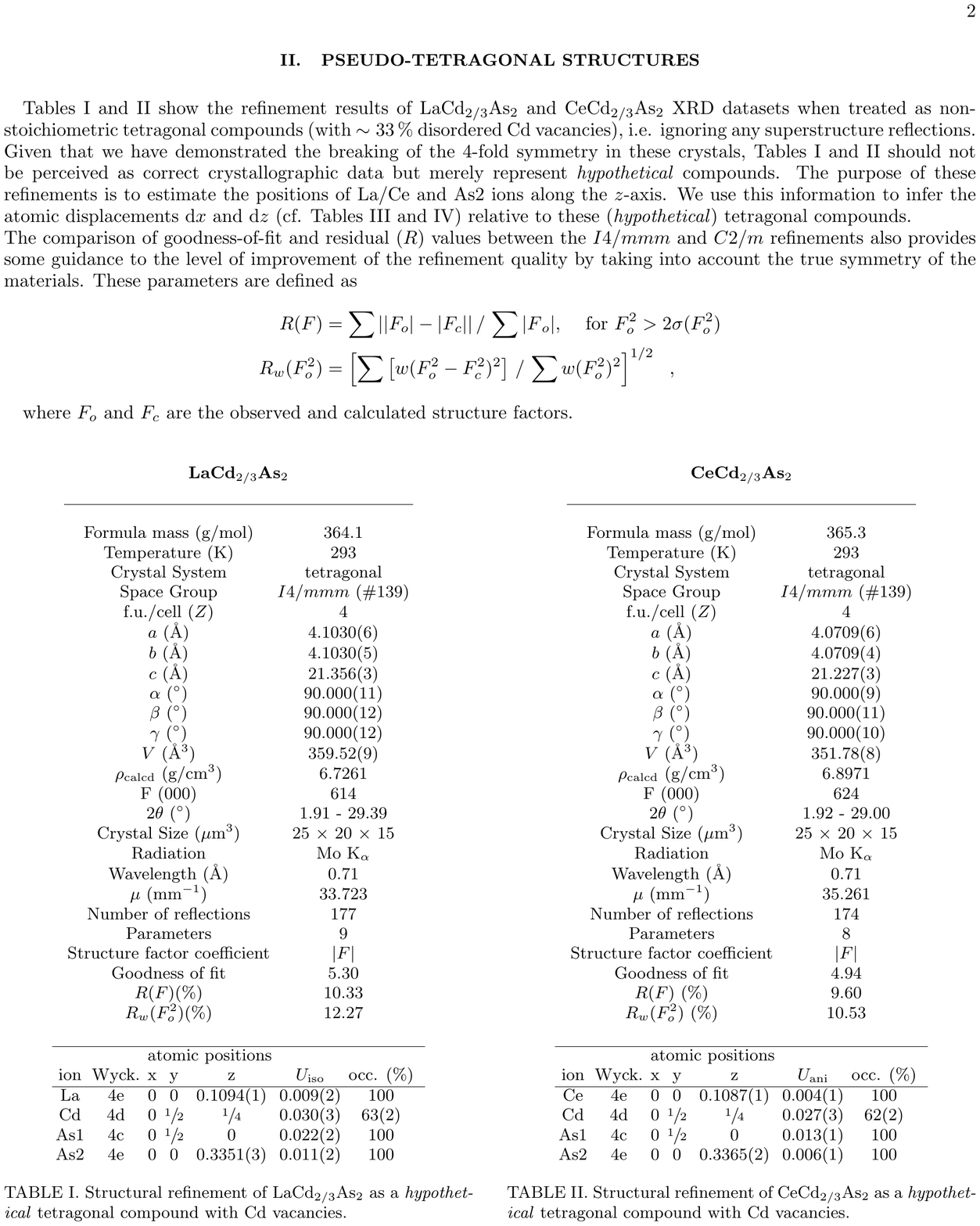}
\end{figure*}

\begin{figure*}[!t]
	\hspace*{-1.2in}
	\vspace*{-1.5in}
	\centering
	\includegraphics[width=1.3\textwidth]{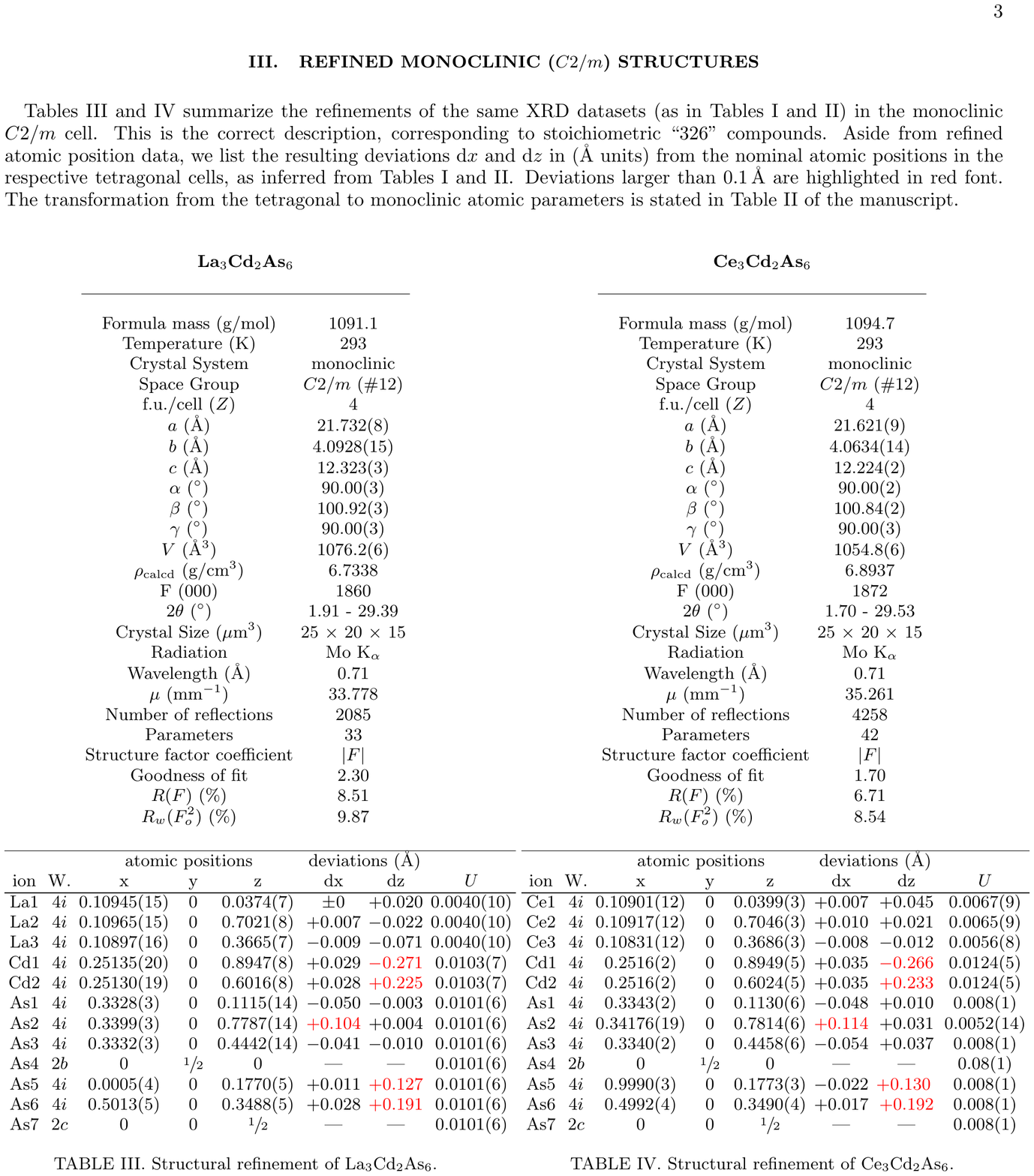}
\end{figure*}

\begin{figure*}[!t]
	\hspace*{-1.2in}
	\vspace*{-1.5in}
	\centering
	\includegraphics[width=1.3\textwidth]{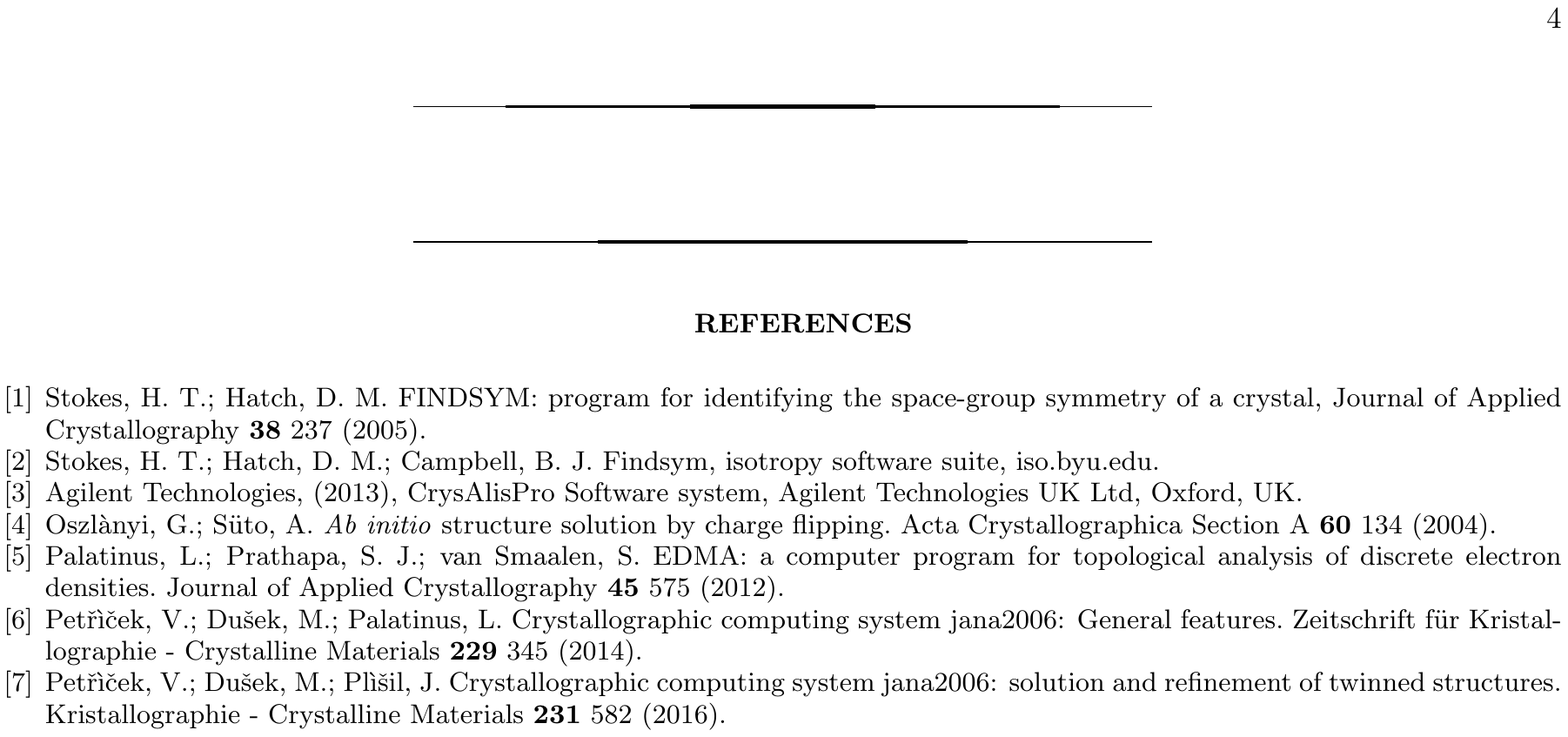}
\end{figure*}
	
\end{document}